\documentclass[fleqn,usenatbib]{mnras}

\usepackage{newtxtext,newtxmath}

\usepackage[T1]{fontenc}

\DeclareRobustCommand{\VAN}[3]{#2}
\let\VANthebibliography\thebibliography
\def\thebibliography{\DeclareRobustCommand{\VAN}[3]{##3}\VANthebibliography}

\usepackage{graphicx}	
\usepackage{amsmath}	
\usepackage{xspace}
\usepackage{longtable}
\usepackage{subfig}
\usepackage{booktabs}
\usepackage{lscape}
\usepackage{siunitx}
\usepackage[symbol]{footmisc}
\usepackage[dvipsnames]{xcolor}
\newcommand{\oxfordastro}{Department of Astrophysics, University of Oxford, Denys Wilkinson Building, Keble Road, Oxford OX1 3RH, UK}
\newcommand{\oxfordmag}{Magdalen College, University of Oxford, Oxford OX1 4AU, UK}

\newcommand{\review}[1]{\textcolor{black}{#1}}

\title[Saturns are not large Neptunes]{Saturns Are Not Large Neptunes: The effect of removing inflated giants from empirical mass-radius relations}

\author[G. Dransfield et al.]{
G. Dransfield,$^{1,2}$\thanks{E-mail: george.dransfield@magd.ox.ac.uk)}
V Tardugno,$^{1}$
Niamh K. O'Sullivan, $^{1}$
\\
$^{1}$ \oxfordastro \\
$^{2}$ \oxfordmag \\
}

\date{Accepted 2026 September 14. Received 2026 August 24; in original form 2026 July 29}

\pubyear{\the\year{}}

\begin{document}
\label{firstpage}
\pagerange{\pageref{firstpage}--\pageref{lastpage}}
\maketitle

\begin{abstract}
Empirical mass-radius relations have long been used in exoplanetology to study planet demographics and to predict masses and radii for objects where only one of these is known. Inferred transitions between planet classes can also provide empirical evidence in support of theory. However, mass-radius relations fitted in two-dimensions can be complicated when additional factors influence the planetary parameters, as is the case with the radius inflation of giant planets. In this work, we present an empirical mass-radius relation derived by fitting a broken power-law to a cleaned PlanetS sample from which inflated giants have been removed. We find that when inflated giants are excluded, the preferred model contains three breakpoints and four segments, showing the emergence of a distinct Saturnian regime between $54\pm3~{\rm M_\oplus} <M< 258\pm11~{\rm M_\oplus}$. In this regime, planets are still growing but at a lower rate than the Neptunian regime, consistent with the onset of gravitational self-compression. We also update the irradiation threshold for giant planet inflation, $S_{\rm thr}=99~\rm S_\oplus$ ($1.3~\rm ergs\,s^{-1}\,cm^{-2}$); below $S_{\rm thr}$,  90\% of giant planets have a radius excess $\Delta \log{R}<2\sigma$.
\end{abstract}




\section{Introduction}

There is a volume mismatch between the outputs of transit and radial velocity surveys: while transit surveys can observe thousands of stars at a time, radial velocities are limited to just one. For this reason, while exceptionally successful space-based transit missions like \textit{Kepler} \citep{Borucki2010} and \textit{TESS} \citep{Ricker2015} have produced $17\,495$ candidate exoplanets between them to date, only 702 have had their masses measured using precise radial velocities\footnote{According to the NASA Exoplanet Archive \citep{Christiansen2025}, accessed on 2026\,July\,02.}. 

Mass-radius relations have a crucial role to play in selecting transiting planets for spectroscopic observations, as they provide an evidence-based approach to predict a planet's mass given its radius. These mass estimates can then be used for semi-amplitude prediction and feasibility assessments for radial velocity follow-ups. Hence, over the years, much work has been dedicated to refining, revising, and revisiting the empirical mass-radius relations of exoplanets \citep[e.g.][]{Seager2007, Fortney2007, Chabrier2009, Chen2017, Parviainen2024, Muller2024}.

While many works use a broken power law to describe the mass-radius relation of exoplanets in two dimensions \citep[e.g.][]{Wolfgang2016, Chen2017, Otegi2020, Muller2024}, alternative non-parametric methods have also been implemented in recent years \citep[e.g.][]{Otegi2020,Parviainen2024}. 

Irrespective of the approach, a further outcome of these exercises is that the exoplanet population is segmented into sub-populations, which also makes mass-radius relations a useful way to study exoplanet demographics. For instance, the forecasting model presented in \cite{Chen2017} not only predicts a planet's mass or radius from the provided parameter, but will also give a predicted category: Terran, Neptunian, Jovian, or star. Other models, such as \cite{Mousavi2023} find only two planet categories: small and giant.

Some models have attempted to include additional parameters such as temperature, stellar irradiation, or stellar mass, which are known to influence the masses and/or the radii of exoplanets \citep[e.g.][]{Edmondson2023, Kanodia2023}. Stellar irradiation is particularly important for giant planets, as highly irradiated giants are known to display significant radius inflation \citep[for a review,][for a review]{Thorngren2024}. The exact mechanism for radius inflation is unknown, but the extent of the effect is known to depend on planetary mass: more massive objects are harder to inflate \citep{Fortney2007}. 

The presence of inflated giants in a sample used to measure empirical mass-radius relations is a complicating factor -- their exquisitely large signals in both transit and RV produce outstanding precision on their mass and radius measurements, which in turn biases the sample. This is especially the case when a sample has been compiled based on precision thresholds \citep[e.g.][]{Otegi2020,Parviainen2024,Muller2024}. \cite{Thorngren2019} circumvented this complication by presenting a mass-radius relation for cool giant exoplanets only, by simply removing all giant exoplanets with $T_{\rm eq}>1000~\rm K$. However, as inflation also depends on mass, some non-inflated objects above this temperature threshold may have been unduly removed.





In this letter, we present for the first time an empirical mass-radius relation spanning the full exoplanet mass range, but with the exclusion of inflated gas giants.

Our paper is structured as follows: in Section \ref{sec:infl} we describe the method for removing inflated giants from our sample. This is followed by Section \ref{sec:mr_fits} where we outline our fitting method for the mass-radius relation, and Section \ref{sec:results} where we present our results in context. Finally, we conclude in Section \ref{sec:concs}.

\section{Removing inflated giants from the sample}
\label{sec:infl}

As mentioned above, \cite{Thorngren2019} filtered out inflated giants by excluding planets with masses $15~{\rm M_\oplus}<M<12~{\rm M_J}$ and equilibrium temperatures $>1000~\rm K$. Another often quoted threshold comes from \cite{Demory2011}\footnote{Although it was first proposed in \cite{Miller2011}.}, where they noted that in their sample of 70 Kepler validated planets, there was no evidence of radius inflation below $2\times 10^8~\rm erg~s^{-1}~cm^{-2}$ ($147~S_\oplus$). However, of the sample presented in that work, only 48 are now classified as confirmed planets;  the remainder have been reclassified as false positives or remain unconfirmed planetary candidates. We will therefore seek to to measure a new stellar irradiation inflation boundary in this work using only confirmed planets.


In this work we use the PlanetS catalogue \citep{Otegi2020}. The benefit of this catalogue is that it only includes planets with precisely measured masses and radii (8\% and 25\% precision respectively); the complete catalogue at the time we accessed it contained 825 planets \footnote{Accessed in May\,2026.}. 

\begin{figure}
    \centering
    \includegraphics[width=1\linewidth]{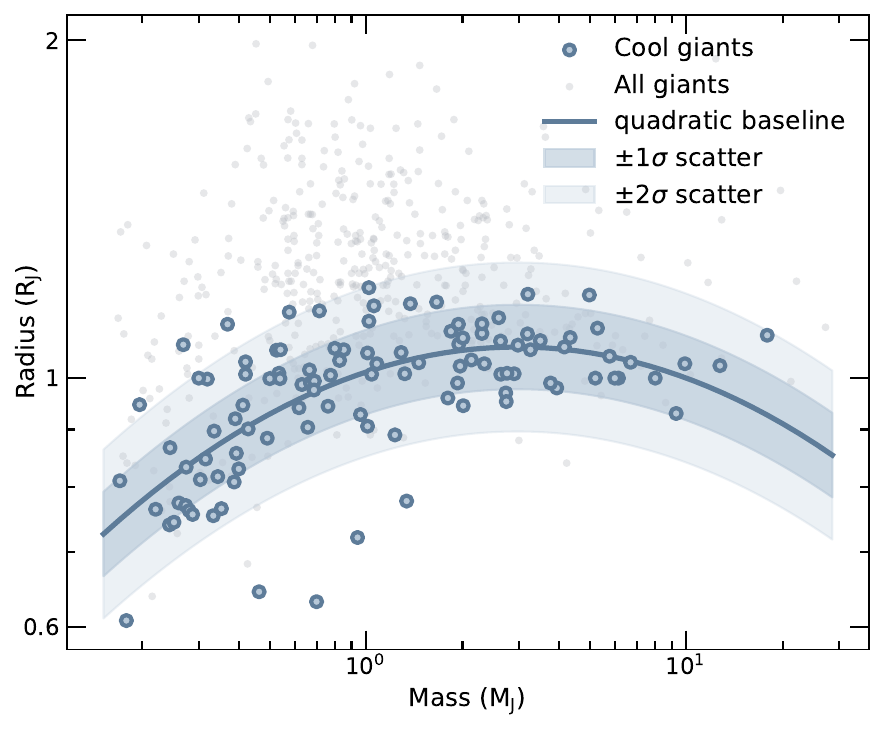}
    \caption{Baseline mass-radius relation fitted to the least irradiated 20\% of giant exoplanets. Blue circles are the sample of cool giants, while the grey points in the background are the remaining giant planet sample. The blue line is the fitted quadratic model, and the blue bands represent the $1\sigma$ and $2\sigma$ scatter. }
    \label{fig:baseline}
\end{figure}

To identify planets inflated due to irradiation, we first construct a subsample of cool giants. We define giant planets as having $R>0.6~\rm R_J$ and $M>0.15~\rm M_J$. These limits were chosen empirically to retain some of the low-mass giant population while excluding lower-mass planets whose radii are dominated by different physical processes, such as inflated hot Neptunes \citep[e.g.][]{Weldon2026} and super-puffs \citep[e.g.][]{Masuda2014}. 
We then rank the sample by incident stellar flux and select the 20\% least irradiated objects. This results in a sample of 108 cool giants, with a maximum stellar irradiation of $S=133.08~\rm S_\oplus$. For the resulting cool giant subset, we fit a mass-radius relation in log space using {\sc sklearn}'s HuberRegressor, a linear regression model that is more robust to outliers than traditional Least-Squares regression. We consider three baseline models - linear, quadratic, and spline - and employ leave-one-out cross-validation (LOO CV) to evaluate their performance. In LOO CV, each planet is excluded from the training set in turn; the model is then fit to the remaining planets, and the $\log R$ of the held-out planet is predicted. Repeating this process for all planets, we sum the absolute differences between the predicted and true log-radii. This procedure provides an estimate of the model's predictive performance on unseen data while minimising overfitting. \review{We find that the quadratic model yields the lowest prediction errors ($\rm LOO\,CV_{RMSE} = 0.0496~dex$), closely followed by the spline model ($\rm LOO\,CV_{RMSE} = 0.0500~dex$), while the linear model performed substantially worse ($\rm LOO\,CV_{RMSE} = 0.0532~dex$). We therefore adopt the quadratic model as the baseline relation for cool giant planets.} The best fit parameters for the squared and linear terms and y-intercept are $-0.10\pm0.02$, $0.10\pm0.01$, and $0.005\pm0.006$ , respectively, and the uncertainties are estimated from the robust covariance matrix of the Huber regression. The fit is shown in Figure \ref{fig:baseline}. 

To estimate the scatter about the baseline relation, we calculate the median absolute deviation (MAD) of the difference (in log-space) between a planet's observed radius and the radius expected from the cool giant baseline model,  $\Delta \log R$, and define the intrinsic scatter as 
\begin{equation}
    \sigma_{\text{int}} = 1.4826 \times \Delta \log R .
\end{equation}

The 1.4826 scale factor converts the MAD into an equivalent standard deviation for a Gaussian distribution. This scatter was used to assess the significance of the logarithmic radius residuals. Planets with $\Delta \log R > N \sigma$, \review{where $N$ is the adopted radius-excess threshold ($N=1,\,1.5\,\rm or\,2$),} were considered larger than expected for their given mass. 

\begin{figure}
    \centering
    \includegraphics[width=1\linewidth]{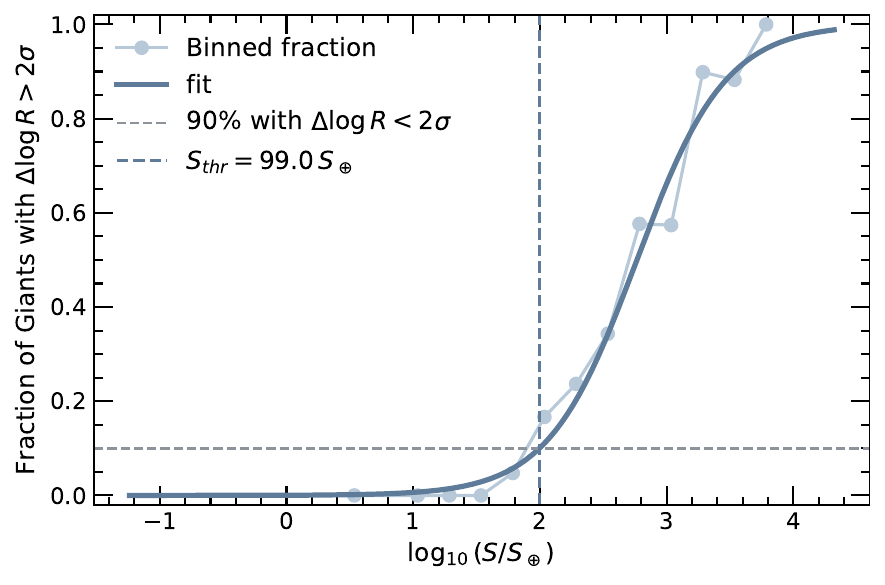}
    \caption{Inflation probability versus insolation. We show the binned fraction of giant planets with $\Delta \log{R}>2\sigma$ in light blue, and the GAM fit in dark blue. The horizontal dashes line indicated the threshold below which 90\% of giant planets have a $\Delta \log{R}<2\sigma$, and the vertical dashed line is where we define our irradiation threshold, $S_{\text{thr}}$.}
    \label{fig:sthr}
\end{figure}

However, a statistically significant radius excess does not necessarily imply inflation. Thus, we also impose an irradiation threshold $S_{\text{thr}}$. Using a logistic generalised additive model (GAM), we fit the fraction of strongly inflated planets ($\Delta \log R > 2 \sigma$) as a function of incident stellar flux. \review{We choose $2\sigma$ as a conservative threshold for strong inflation as these planets are significantly larger than expected based on the cool giant mass-radius relation.} We then define the threshold as the irradiation at which the model predicts a 10\% probability of strong inflation. A planet is therefore considered inflated if it both has a radius residual greater than \review{the adopted} $N\sigma$ \review{threshold} and its irradiation exceeds $S_{\text{thr}}$. We find $S_{\rm thr} = 99.02 S_\oplus$ (See Figure \ref{fig:sthr}). \review{This updated physical irradiation threshold for the onset of strong giant planet inflation is held fixed for the remainder of our analysis.}


\section{Fitting the mass-radius relation}
\label{sec:mr_fits}

Following the inflation correction described in Section~\ref{sec:infl}, we define
\begin{equation}
x=\log_{10}\left(\frac{M}{M_\oplus}\right)
\qquad {\rm and} \qquad
y=\log_{10}\left(\frac{R}{R_\oplus}\right),
\end{equation}
and assume that the mass-radius relation can be described by a broken power law with an unknown number of breakpoints:
\begin{equation}
\label{eq:mr_model}
    y = c + ax + \sum_{j=1}^{k}d_j\,\max\left(0,x-\psi_j \right) \, , 
\end{equation}
where $c$ is the intercept, $a$ is the slope of the first segment, $d_j$ are slope changes, $\psi_j$ are breakpoint locations, and $k$ is the number of breakpoints, which is a free parameter in our fit. This parameterisation ensures continuity of the mass-radius relation across all breakpoints. As the first slope is $m_1 = a$, subsequent slopes following breakpoints are defined as
\begin{equation}
\label{eq:slopes}
    m_i = a + \sum_{j=1}^{i-1} d_j \, ,
\end{equation}
such that each breakpoint modifies the slope of the relation by an amount $d_j$.

As both mass and radius measurements have uncertainties we must take into account, we fit our model using orthogonal distance regression (ODR) \citep{Boggs1987} which accounts for uncertainties in both variables simultaneously. We implement this via the {\sc Python} package {\sc odrpack}. We fitted models containing between 0 and 4 breakpoints and calculated the Bayesian Information Criterion (BIC), which balances goodness-of-fit with model complexity. We then adopted the model with the lowest BIC. 

We estimated the uncertainties on our parameters using the covariance matrix returned by the ODR solution. Assuming the parameter uncertainties are locally Gaussian, we used the covariance matrix returned by the ODR solution to construct a multivariate normal distribution centred on the best-fit parameter vector. We generated $2\times 10^4$ random samples of the model parameters from this distribution and transformed them into the corresponding breakpoint locations and segment slopes using Equations \ref{eq:mr_model} and \ref{eq:slopes}. We then estimated parameter uncertainties from the 16$^{\rm th}$, 50$^{\rm th}$ and 84$^{\rm th}$ percentiles of the resulting distributions. 

We fitted our mass-radius model to nine versions of the inflation-cut sample. For each of the three baseline models fitted in Section \ref{sec:infl}, we test three radius excess cut-offs: $1\sigma$ (mild inflation), $1.5\sigma$ (moderate inflation), $2\sigma$ (strong inflation). Our tests in Section \ref{sec:infl} revealed that the quadratic model had the lowest intrinsic scatter and the highest predictive power for unseen planets, and thus this is the model we ultimately adopt as our fiducial inflation model. The purpose of these nine combinations is therefore not to optimise the inflation cut, but rather to assess the robustness of recovered breakpoints and slopes to choices of baseline model and inflation threshold. We thus adopt as a final mass-radius model the one which best combines statistical performance with stability of the inferred planetary regimes across alternative sample definitions.

\section{Results and discussion}
\label{sec:results}

\begin{figure*}
    \centering
    \includegraphics[width=0.9\textwidth]{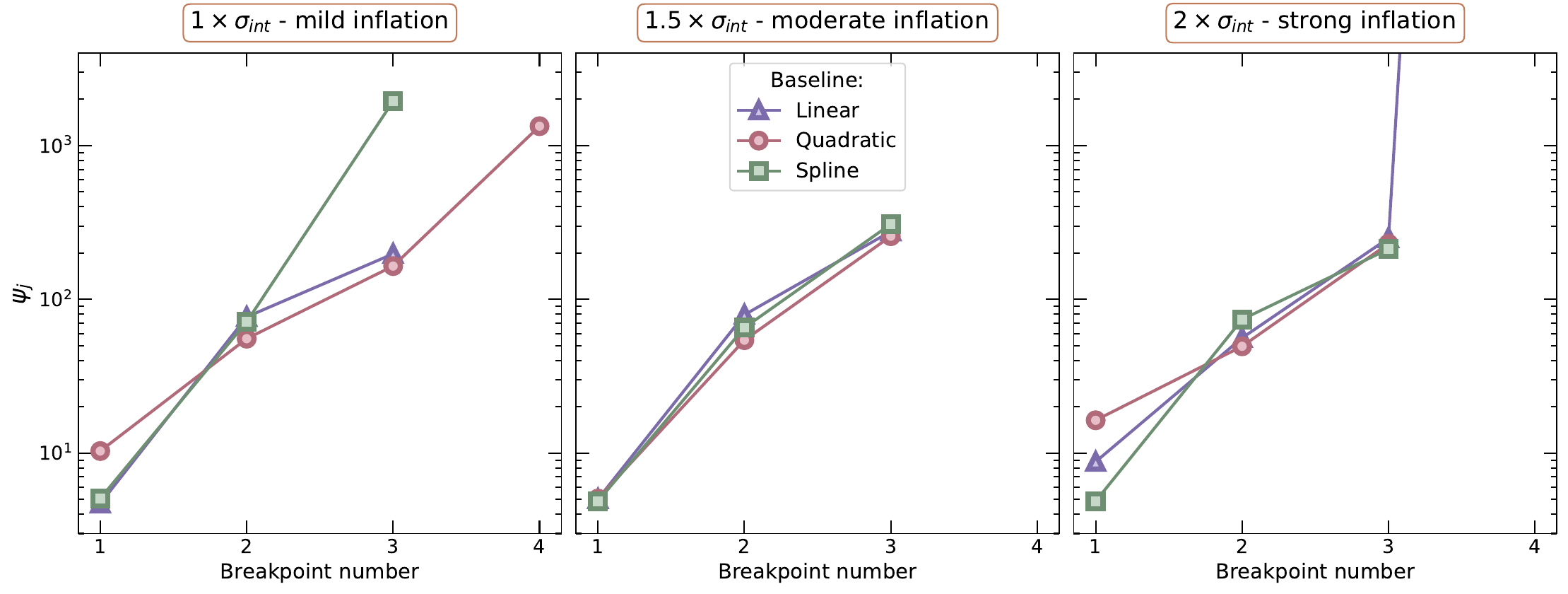}
    \includegraphics[width=0.9\textwidth]{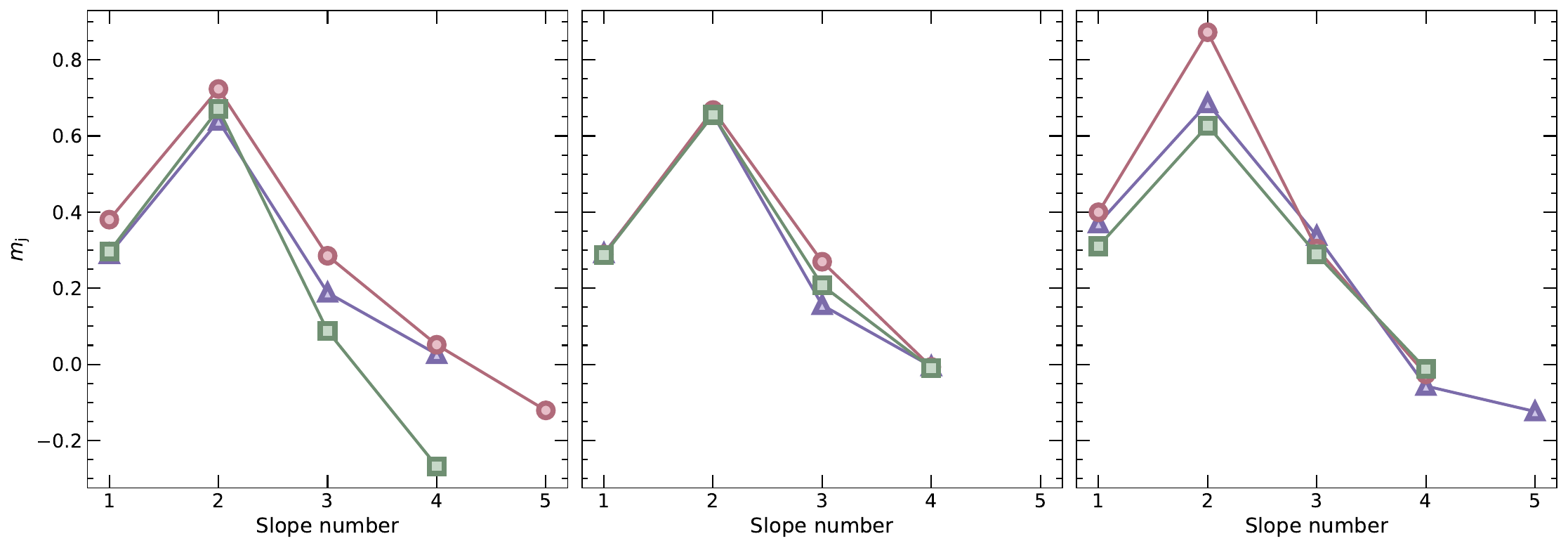}
    \caption{Breakpoints and segment slopes fitted for the nine combinations of baseline model and inflation threshold. The top row presents the fitted breakpoints for each inflation threshold, and the bottom presents the segment slopes. We note that in the upper right panel, the final fitted breakpoint was well beyond the range of the data, and we therefore don't display it on the figure. }
    \label{fig:all_fits}
\end{figure*}

In Figure \ref{fig:all_fits} we present the fitted breakpoints and segment slopes for all nine combinations of baseline model and inflation threshold. All nine fits identify a breakpoint between Neptune and Saturn, while eight out of nine also identify a breakpoint between Saturn and Jupiter. These results indicate that the observed mass-radius relation strongly favours the existence of a distinct Saturnian regime, regardless of the adopted inflation correction or baseline model. 

In all but two of the fits (linear, $2\sigma$ and quadratic, $1\sigma$), the number of breakpoints with the lowest BIC was $k=3$; the remaining two preferred the inclusion of a fourth breakpoint. 

\begin{figure}
    \centering
    \includegraphics[width=\columnwidth]{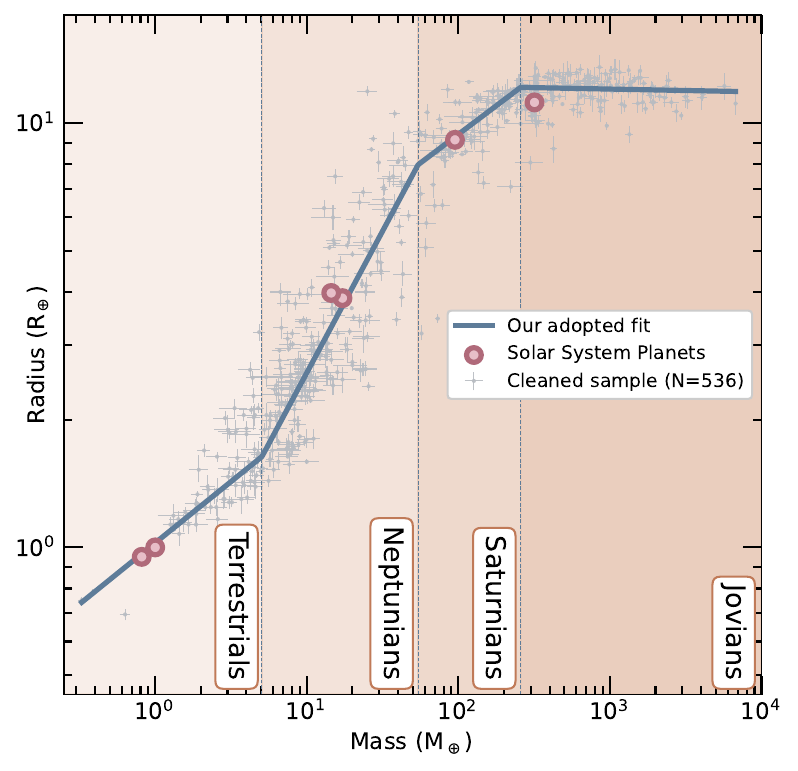}
    \caption{Our adopted mass-radius relation. The blue line shows our fitted piecewise model and the grey points in the background are the 536 planets in our cleaned sample. We overplot relevant solar system planets in pink, and highlight the positions of the breakpoints which mark the boundaries between the different labelled planetary regimes. }
    \label{fig:final_fit}
\end{figure}

\subsection{Description of results and comparison to previous works}

We find that the fit to the sample with a moderate inflation cut ($1.5\sigma$) show the most stability in fitted breakpoints and slopes across the three baseline definitions (see Figure \ref{fig:all_fits}, middle column). \review{In all three baseline cases the model contained three breakpoints, whereas the $1\sigma$ and $2\sigma$ samples each produced one baseline model that favoured a fourth breakpoint. Additionally, the recovered breakpoint locations and segment slopes showed the smallest variation across the three baseline models for the $1.5\sigma$ sample; in particular, the terrestrial breakpoint was consistently found at $4.9-5.1~\rm M_{\oplus}$, the Neptunian-Saturnian breakpoint at $54-79~\rm M_{\oplus}$, and the Jovian slope is consistent with zero in all three fits. We therefore adopt as our final mass-radius relation the model fitted to the quadratic, $1.5\sigma$ sample.} The adopted fit \review{and the cleaned sample are} presented in Figure \ref{fig:final_fit}; \review{the cleaned sample consists of the PlanetS catalogue after excluding inflated giants using the procedures detailed in Section \ref{sec:infl}}. The parameters for our mass-radius relation are as follows:
\begin{equation}
\label{eq:mr_final}
R =
\begin{cases}
(1.02 \pm 0.01)\,M^{0.289\pm0.006} &
M < (5.1 \pm 0.2) \\
\\
(0.554 \pm 0.010)\,M^{0.668\pm0.007} &
(5.1 \pm 0.2) < M  \\ &< (54.3 \pm 2.7) \\
\\
(2.72 \pm 0.10)\,M^{0.270\pm0.007} &
(54.3 \pm 2.7) < M \\ & < (257.8 \pm 10.8) \\
\\
(12.6 \pm 0.3)\,M^{-0.007\pm0.004} &
M > (257.8 \pm 10.8)
\end{cases}
\end{equation}
Our fit with three breakpoints separates the planet population into four distinct regimes which we label as Terrestrials, Neptunians, Saturnians, and Jovians. We find the transition from the Terrestrial to the Neptunian regime to be at $M=5.1 \pm 0.2~\rm M_{\oplus}$, which is consistent with the transition at $M=4.37 \pm 0.72~\rm M_{\oplus}$ found by \cite{Muller2024}, although the transition is expected to be at $10~\rm M_\oplus$ based on early theoretical works \citep[e.g.][]{Seager2007, Fortney2007}. The corresponding transition in radius is $R=1.63\pm0.07~\rm R_\oplus$, consistent with the expectation from \cite{Rogers2015, Cloutier2020, Parviainen2024} that planets with $R<1.6~\rm R_\oplus$ are rocky. However, our fitted breakpoint represents only the average transition between the Terrestrial and Neptunian regimes, and cannot capture the known compositional degeneracy in the small-planet population. As evident in Figure \ref{fig:final_fit}, planets with masses between $\sim 2-10~\rm M_{\oplus}$ occupy two approximately parallel sequences corresponding to rocky and volatile-rich worlds \citep{Luque2022}, resulting in substantially larger scatter about the fitted relation than in the other regimes. 
The slope of the Terrestrial planet regime in our fit is $0.289\pm0.006$, which is comparable to theoretical models that predict $R \propto M^{1/3}$ for solid spheres. Our result is also compatible with the slopes of $0.28\pm0.01$, $0.29\pm0.01$, and $0.27\pm0.04$ found by \cite{Chen2017, Otegi2020} and \cite{Muller2024} respectively. One key difference with the results of \cite{Otegi2020}, however, is that their piecewise model separates the rocky and volatile-rich populations by density, allowing planets with the same masses to belong to different regimes.

Our Neptunian regime has a fitted slope of $0.668\pm0.007$. In this regime we expect to find volatile-rich planets which have accreted H-He envelopes, and planets see the most rapid growth with the addition of more mass. The slope we find is consistent once again with \cite{Otegi2020} and \cite{Muller2024}, who find slopes of $0.63\pm0.04$ and $0.67\pm0.05$ respectively; however, \cite{Chen2017} find a slightly lower Neptunian slope of $0.59\pm0.04$, likely attributable to the sparser planet sample available at the time. 

The key difference between our mass-radius relation and previous works comes with the introduction of a breakpoint separating a Saturnian regime from the Neptunian one. We find this breakpoint to be at $M=54.3\pm2.7~\rm M_\oplus$ with a corresponding radius transition at $R=8.0\pm0.5~\rm R_\oplus$, leading into a Saturnian slope of $0.270\pm0.007$. The Saturnian slope is substantially lower than the Neptunian slope yet it remains decidedly positive, suggesting that the rapid growth in radius associated with increasing envelope mass is already being moderated by gravitational self-compression. However, as radius growth is still evident, we can see that envelope growth still dominates over compressional effects. The observed onset of self-compression in the Saturnian regime is is consistent with the increasing importance of electron degeneracy in hydrogen-rich interiors, which \cite{Chabrier2009} notes begins around giant-planet masses, and not at the hydrogen burning limit. 

The final transition is from Saturnian planets onto Jovians, which we find takes place at a mass of $M=257.8\pm10.8~\rm M_\oplus$ and a radius of $R=12.2\pm0.5~\rm R_\oplus$, and the slope of the Jovian regime is almost flat: $-0.007\pm0.004$. This is consistent with previous results and with theoretical expectations, as here electron degeneracy pressure prevents further growth with the addition of more mass, resulting in radii that are independent of mass \citep{Burrows2001, Chabrier2009}. The existence of two distinct giant planet regimes is also consistent with the onset of degeneracy being gradual, rather than occurring at one single critical mass -- this then results in a gradual flattening of the mass-radius relation.

\cite{Helled2023} recently posited that the transition to H-He dominated gas giants may occur at masses $\sim 100\rm ~ M_\oplus$, suggesting that Saturn never underwent runaway gas accretion, and that it therefore occupies a distinct structural regime. Our results provide independent empirical support for this interpretation. The Saturnian branch identified by our model spans $54.3\pm2.7$ to $257.8\pm10.8$ with a mass-radius slope intermediate between the growing Neptunian regime and the nearly flat Jovian regime. This behaviour is consistent with a population of planets that have gained substantial H-He enveloped but have not yet reached the strongly self-compressed state characteristic of Jovian planets. 

\subsection{Is a fourth breakpoint physically plausible?}

We noted above that in two of the nine fits, the lowest BIC was found for the fit with four breakpoints. We disregard the linear, $2\sigma$ fit as the fourth breakpoint was well beyond the boundary of the dataset, indicating an unstable fit. However we discuss here whether the five-segment fit resulting from the quadratic, $1\sigma$ fit is physically plausible. 

Similar to previous works, we used the full available mass range present in the PlanetS sample; this includes objects with masses $M>13~M_{J}$ which is the minimum mass to ignite deuterium fusion. This should not be considered a hard and fast transition from planets to brown dwarfs, however, as the manner and environment in which objects formed is a more important predictor of their nature \citep{Burrows2001}. Nevertheless, \cite{Chabrier2009} notes that as masses increase into the brown dwarf regime and electron degeneracy becomes more important, the expected slope of the mass radius relation is $-1/8=-0.125$. This is consistent with self-compression having a larger effect, leading to radii $\sim0.9~\rm R_J$ for most brown dwarfs. 

The slope for the fifth segment found by the quadratic, $1\times\sigma$ fit is $m_5=-0.121\pm0.019$ with a breakpoint at $\psi_4=1341.1\pm82.5 \rm ~M_{\oplus}=4.2\pm0.3~M_{J}$. The fitted slope is remarkably consistent with theoretical expectations for increasingly degenerate hydrogen-rich objects \citep{Chabrier2009}, suggesting the model may be identifying the transition from Jovian planets to the brown dwarf regime. However, the inferred transition mass is substantially lower than the conventional deuterium burning limit of $13~\rm M_J$, and the existence of a fifth segment is not robust across all our fits. We also note that other works place the giant planet-brown dwarf boundary at even higher masses, with \cite{Schneider2011} adopting $25~\rm M_J$, and \cite{Hatzes2015} finding that giant planets extend to $\sim 60~\rm M_J$.

At present, the sample contains relatively few objects in the brown dwarf mass regime, which significantly limits our ability to constrain the mass-radius relation beyond giant planets. We therefore do not consider the evidence for a distinct brown dwarf branch to be sufficiently compelling at present. \review{Additionally, the fact that this branch is only recovered in the most aggressively cleaned sample ($1\sigma$) suggests that it may be a consequence of over-pruning the giant planet population, rather than evidence for an additional physically distinct regime.} Nevertheless, the excellent agreement of the slope with theoretical expectations makes this a compelling avenue for future study as the sample of precisely characterised brown dwarfs grows. 

\subsection{Uncertainty Estimation}

We report the results of our best fit model along with uncertainties estimated from the covariance matrix of the orthogonal distance regression. These uncertainties are likely to underestimate the parameter uncertainties since they rely on local approximations of the parameter space. In principle, for a non-linear, non-smooth model such as the mass-radius relation we are fitting, bootstrap resampling would provide more robust uncertainty estimates. However, the relatively small dataset limits the reliability of resampling techniques for uncertainty estimation, especially techniques requiring the removal of many data points at a time. An alternative resampling method to approximate uncertainties is jackknife resampling. The procedure consists of removing one data point at a time, fitting the model to the remaining sample, and recomputing all statistics of interest. For each statistic $\theta$, the jackknife mean can be computed as 
\begin{equation}
    \bar \theta = \frac{1}{N} \sum_i^N \theta_i
\end{equation}
where $\theta_i$ corresponds to the calculated statistic for the $i$th jackknife realisation. The median leave-one-out deviation is then

\begin{equation}
    \Delta_{\text{LOO}}
= \sqrt{(N-1 ) \text{median}_i [
(\theta_i-\bar{\theta})^2]}.
\end{equation}

\begin{figure}
    \centering
    \includegraphics[width=\columnwidth]{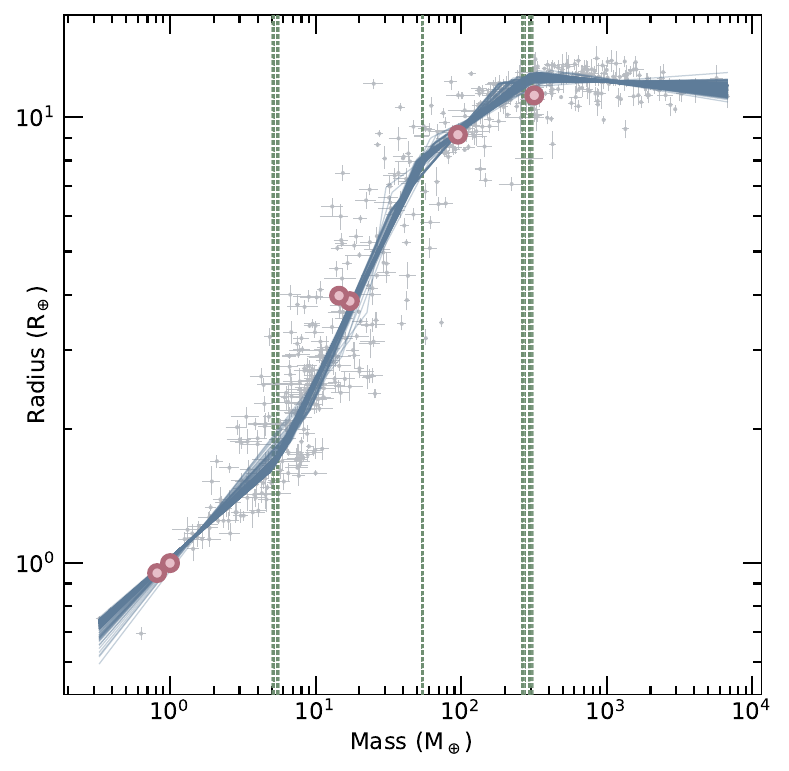}
    \caption{Mass-radius relation computed for jackknife realisations with breakpoints between the 16th and 84th percentiles. }
    \label{fig:jackknife}
\end{figure}

Jackknife resampling results in relatively stable slopes and overall mass-radius relation (see Figure \ref{fig:jackknife}), with uncertainties of $\pm$ 0.126, 0.217, 0.124, and 0.189 for the four slopes. On the other hand, the distributions of the estimated breakpoints are strongly concentrated around the best fit values, but have a long tail resulting from a small number of observations around the transition points. The omission of these influential observations can have a significant effect on the threshold and breakpoint fits, resulting in large variations in the inferred parameters and the uncertainties being dominated by this small subsample. Despite most jackknife realisations yielding breakpoint values closely centred around the best-fit values with standard deviations of 1.28, 4.00, and 37.04, the aforementioned tail results in jackknife deviations of $\pm$  6.46, 11.84, and 397.36. 

Moreover, jackknife operates under the assumption that the estimator is a smooth functional of the data. This assumption is not satisfied by piecewise models with breakpoints, such as the one we report. Thus, we interpret the jackknife results qualitatively, which indicate a robustness of the overall mass-radius relation while the precise locations of the breakpoints are less well constrained.


\section{Conclusions}
\label{sec:concs}

In this paper we have presented an empirical mass-radius relation fitted to a sub-sample of the PlanetS catalogue, where inflated giant planets have been excluded. Our main results are as follows:

\begin{itemize}
    \item We update the well-known irradiation threshold for giant planet inflation presented in \cite{Demory2011}, defining $S_{\rm thr}=99~\rm S_\oplus$, the irradiation below which 90\% of giant planets have a radius excess $\Delta\log{R}<2\sigma$.
    \item We fitted our mass-radius relation to nine versions of the inflation-cleaned sample and find that in all cases, a breakpoint was placed between Neptune and Saturn, while in eight out of nine there was also a breakpoint between Saturn and Jupiter. 
    \item Our best-fit mass-radius relation has three breakpoints and four distinct segments, one of which corresponds to a regime of Saturnian planets. This newly defined regime is between $54.3\pm2.7~{\rm M_\oplus} <M< 257.8\pm10.8~{\rm M_\oplus}$ and has a slope of $0.270\pm0.007$, with a corresponding radius range of $8.0\pm0.5~{\rm R_\oplus} <R< 12.2\pm0.5~{\rm R_\oplus}$. 
    \item The lower slope of the Saturnian regime compared to the Neptunian regime ($0.668\pm0.007$) is consistent with the increased importance of electron degeneracy in hydrogen-rich interiors, and the corresponding onset of gravitational self-compression. 
\end{itemize}

We emphasise that the PlanetS sample is biased by design, as selecting planets based on the precision of mass and radius measurements favours planets with larger masses and radii, and those with shorter orbital periods. In the coming years, as the sample of precisely characterised planets continues to grow across all mass, radius and period regimes thanks to upcoming missions like \textit{PLATO} \citep{plato}, \textit{The Nancy Grace Roman Space Telescope} \citep{roman}, and the Terra Hunting Experiment \citep{the}, it will be crucial to continue fitting new mass-radius relations to improve our understanding of the finer structure of the exoplanet population. 

\section*{Acknowledgements}
\review{The authors would like to thank the anonymous reviewer for their helpful comments, which allowed us to clarify and improve the manuscript. }
GD acknowledges funding from Magdalen College, Oxford. VT acknowledges funding from the Clarendon Fund. NKOS acknowledges funding from the European Research Council under the European Union’s Horizon 2020 research and innovation programme (grant agreement no. 865624, GPRV). The authors would also like to thank Suzanne Aigrain and David Hogg for helpful discussions and suggestions while preparing this manuscript.
This work makes extensive use of {\sc Scipy} \citep{scipy}, {\sc Numpy} \citep{numpy}, {\sc Scikit-learn} \citep{scikit-learn}, {\sc Pandas} \citep{pandas}, and {\sc Matplotlib} \citep{matplotlib}. 

\section*{Data Availability}

The data used in this work are part of the PlanetS catalogue, presented in \cite{Otegi2020}. The catalogue is publicly available at \url{https://dace.unige.ch/exoplanets/?}.



\bibliographystyle{mnras}
\bibliography{example} 




\appendix




\bsp	
\label{lastpage}
\end{document}